\documentclass[10pt, conference, compsocconf]{IEEEtran}
\usepackage{graphicx}
\usepackage{pifont}
\usepackage[usenames,table]{xcolor}
\usepackage{flushend}
\usepackage{url}

\begin{document}
%
\title{Cloud Forensics: A Meta-Study of Challenges, Approaches, and Open Problems}

\author{\IEEEauthorblockN{Shams Zawoad}
\IEEEauthorblockA{University of Alabama at Birmingham\\
Birmingham, Alabama 35294-1170\\
Email: zawoad@cis.uab.edu}
\and
\IEEEauthorblockN{Ragib Hasan}
\IEEEauthorblockA{University of Alabama at Birmingham\\
Birmingham, Alabama 35294-1170\\
Email: ragib@cis.uab.edu}}

\maketitle

\begin{abstract}
In recent years, cloud computing has become popular as a cost-effective and efficient computing paradigm. Unfortunately, today's cloud computing
architectures are not designed for security and forensics. To date, very little research has been done to develop the theory and practice of cloud forensics.
Many factors complicate forensic investigations in a cloud environment. First, the storage system is no longer local. Therefore, even with a subpoena, law enforcement agents cannot confiscate the suspect's computer and get access to the suspect's files. Second, each cloud server contains files from many users. Hence, it is not feasible to seize servers from a data center without violating the privacy of many other users. Third, even if the data belonging to a particular suspect is identified, separating it from other users' data is difficult. Moreover, other than the cloud provider's word, there is usually no evidence that links a given data file to a particular suspect. For such challenges, clouds cannot be used to store healthcare, business, or national security related data, which require audit and regulatory compliance.

In this paper, we systematically examine the cloud forensics problem and explore the challenges and issues in cloud forensics. We then discuss existing research projects and finally, we highlight the open problems and future directions in cloud forensics research area. We posit that our systematic approach towards understanding the nature and challenges of cloud forensics will allow us to examine possible secure solution approaches, leading to increased trust on and adoption of cloud computing, especially in business, healthcare, and national security. This in turn will lead to lower cost and long-term benefit to our society as a whole.

\end{abstract}
\IEEEpeerreviewmaketitle

\section{Introduction}
\label{sec:introduction}
Cloud computing has emerged as a popular and inexpensive computing paradigm in recent years. In the last 5 years alone, we have seen an explosion of applications of cloud computing technology, for both enterprises and individuals seeking additional computing power and more storage at a low cost. Small and medium scale industries find cloud computing highly cost effective as it replaces the need for costly physical and administrative infrastructure, and  offers the flexible pay-as-you-go structure for payment.  Khajeh-Hosseini et al.  found that an organization could save 37\% cost if they would migrate their IT infrastructure from an outsourced data centre to the Amazon's Cloud \cite{khajeh2010cloud}. A recent research by Market Research Media states that the global cloud computing market is expected to grow at an 30\% Compound Annual Growth Rate (CAGR) reaching \$270 billion in 2020 \cite{websitemarketresearchmedia}. According to Gartner Inc., the strong growth of cloud computing will bring \$148.8 billion revenue by 2014 \cite{websitegartnernews}. Cloud computing is getting popular not only in the private industry, but also in the government sector. According to a research from INPUT, the US Federal government's spending on the cloud will reach \$792 million by 2013 \cite{input2009}.

Clouds use the multi-tenant usage model and virtualization to ensure better utilization of resources. However, these fundamental characteristics of cloud computing are actually a double-edged sword -- the same properties also make cloud-based crimes and attacks on clouds and their users difficult to prevent and investigate.  According to a recent IDCI survey, 74\% of IT executives and CIOs referred security as the main reason to prevent their migration to the cloud services model \cite{websiteclavister}. Some recent attacks on cloud computing platforms strengthen the security concern. For example, a botnet attack on Amazon's cloud infrastructure was reported in 2009 \cite{websiteamazon2009}. Besides attacking cloud infrastructure, adversaries can use the cloud to launch attack on other systems. For example, an adversary can rent hundreds of virtual machines (VM) to launch a Distributed Denial of Service (DDoS) attack. After a successful attack, she can erase all the traces of the attack by turning off the VMs. A criminal can also keep her secret files (e.g., child pornography, terrorist documents) in cloud storage and can destroy all evidence from her local storage to remain clean. To investigate such crimes involving clouds, investigators have to carry out a digital forensic investigation in the cloud environment. This particular branch of forensic has become known as  \emph{Cloud Forensics}. 
	
According to an annual report of Federal Bureau of Investigation (FBI), the size of the average digital forensic case is growing at 35\% per year in the United States. From 2003 to 2007, it increased from 83GB to 277 GB in 2007 \cite{fbi2008}. This rapid increase in digital forensics evidence drove the forensic experts to devise new techniques for digital forensics. At present, there are several established and proven digital forensics tools in the market. With the proliferation of clouds, a large portion of these investigations now involves data stored in or actions performed in a cloud computing system. Unfortunately, many of the assumptions of digital forensics are not valid in cloud computing model. For example, in a cloud environment, investigators do not have physical access to the evidence -- something they usually have in traditional privately owned and locally hosted computing systems. As a result, cloud forensics brings new challenges from both technical and legal point of view and has opened new research area for security and forensics experts. \\
	
\noindent\textbf{Contributions.}~In this article, we present a systematic analysis of the cloud forensics problem. The contributions of this paper are as follows: 
\begin{itemize}
\item We present a systematic summary of the challenges and issues in cloud forensics.
\item We provide a comprehensive analysis of proposed solutions for cloud forensics in the three different service models of publicly deployed cloud computing. 
\item We also identify the usages and advantages of cloud computing in digital forensics and enumerate the current open problems of cloud forensics.\\
\end{itemize}
	
\noindent\textbf{Organization.~}	
The rest of the article is organized as follows: Section \ref{sec:background} provides the background knowledge of  cloud computing, digital forensics, and cloud forensics. Section \ref{sec:challenge} presents the challenges in cloud forensics and section \ref{sec:solution} discusses the existing proposed solutions. Section \ref{sec:evaluation} provides an evaluation of existing digital forensics tools in a cloud environment. In Section \ref{sec:advantage}, we discuss the advantages of cloud forensics over traditional computer forensics and Section \ref{sec:usageindigital} describes some use cases of cloud computing in digital forensics. Section \ref{sec:problem} presents the open problems of cloud forensics and finally, we conclude in Section \ref{sec:conclusion}. 

\section{Background}
\label{sec:background}

In this section, we provide a brief overview of cloud computing and computer forensics. We also discuss the unique nature of clouds, which make digital forensics investigations difficult.

\subsection{Cloud computing }

\noindent\textbf{Definition.~}According to the definition by the  National Institute of Standards and Technology (NIST), \textit{``Cloud computing is a model which provides a convenient way of on-demand network access to a shared pool of configurable computing resources (e.g., networks, servers, storage, applications, and services), that can be rapidly provisioned and released with minimal management effort or service provider interaction''} \cite{mell2009draft}. The Open Cloud Manifesto Consortium defines cloud computing as \textit{``the ability to control the computing power dynamically in a cost-efficient way and the ability of the end user, organization, and IT staff to utilize the most of that power without having to manage the underlying complexity of the technology''} \cite{opencloudmenifesto2009}. 

Cloud computing has some important characteristics -- On-demand self-service, broad network access, resource pooling, rapid elasticity, and measured service. Parkhill proposed utility computing long ago \cite{parkhillchallenge} and Michael et al. mention cloud computing as a new term for ``computing as a utility'' \cite{michael2009above}. They defined cloud computing as a combination of Software-as-a-Service and utility computing, but they consider private clouds outside of cloud computing. \\

\begin{figure*}[!ht]
\centering
\includegraphics[width=.9\textwidth]{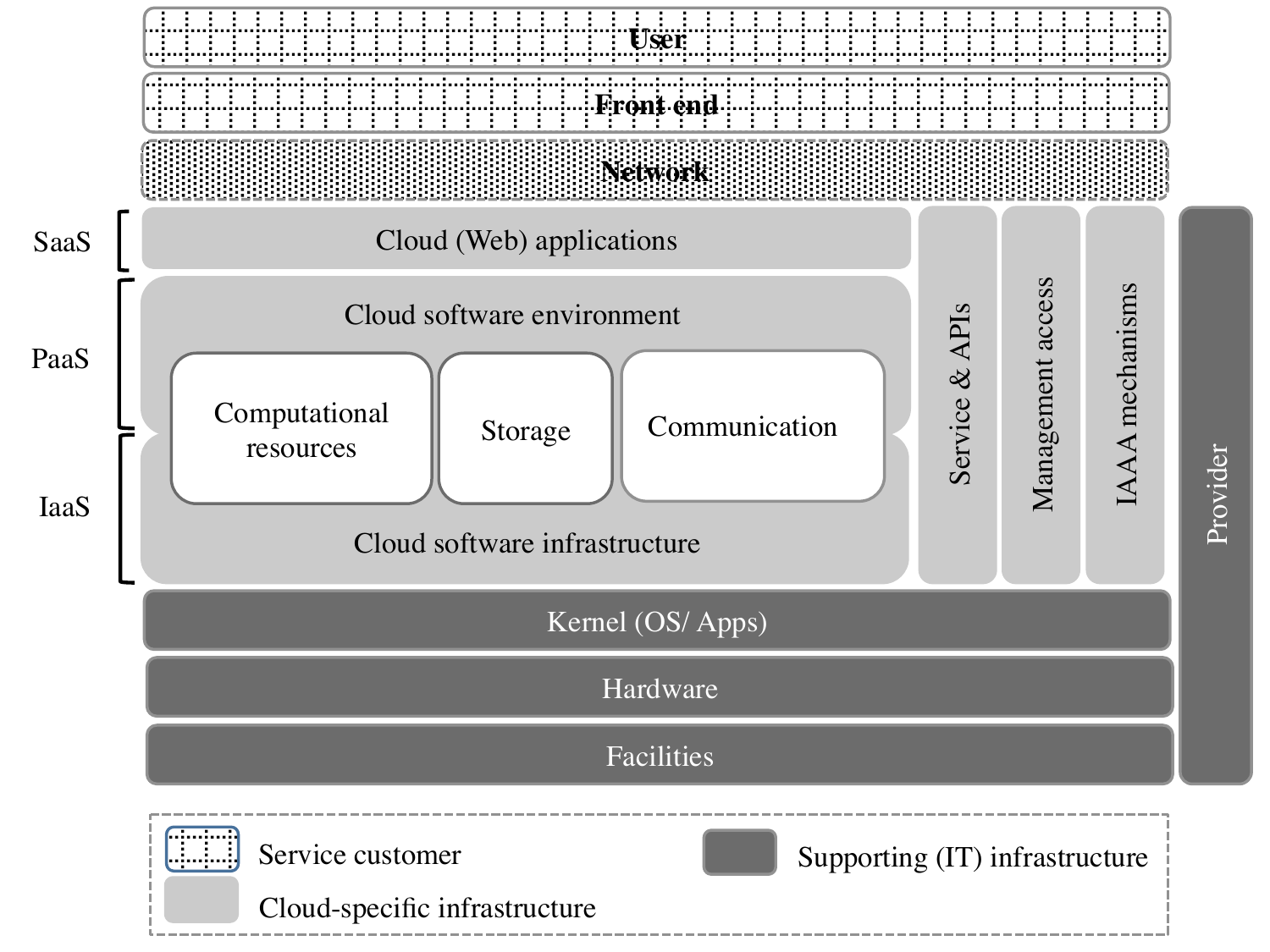}
\caption{Three service models of Cloud Computing \protect\cite{Grobauer:2010:TIH:1866835.1866850}}
\label{figure:cloudservice}
\end{figure*}

\noindent\textbf{Classification according to service model.~}According to the nature of service model used by the Cloud Service Provider (CSP), cloud computing can be divided into three categories: Software as a Service (SaaS), Platform as a Service (PaaS), and Infrastructure as a Service (IaaS) \cite{mell2009draft}. Figure \ref{figure:cloudservice} illustrates the three service models in cloud computing architecture. \\ \smallskip

\noindent \textbf{Software as a Service (SaaS).}~This model provides the consumers the facility of using cloud service provider's software application running on cloud infrastructure. This approach is different from traditional software package distribution to individuals or organizations. In this model, there is no need for software distribution. Consumers can access the application through the web browsers in computers or mobile devices. Usually, there is a monthly subscription fee to use the service. This fee can sometimes vary according to the number of users of an organization. In this model, customers do not have any control over the network, servers, operating systems, storage, or even on the application, except some access control management for multi-user application. Some of the examples of SaaS are : Salesforce \cite{websitesalesforce}, Google Drive \cite{websitegoogledrive}, and Google calender \cite{websitegooglecalendar}. \smallskip

\noindent \textbf{Platform as a Service (PaaS).}~In PaaS, customers can deploy their own application or a SaaS application in the cloud infrastructure. Normally, the customers pay according to the bandwidth usage and database usage. They do not manage or control the underlying cloud infrastructure including network, servers, operating systems, or storage, but have control over the deployed applications and some application hosting environment configurations. Customers can only use the application development environments, which are supported by the PaaS providers. Two examples of PaaS are: Google App Engine (GAE) \cite{websitegae} and Windows Azure \cite{websiteazure}. Customers can host their own developed web based application on these platforms.\smallskip

\noindent \textbf{Infrastructure as a Service (IaaS).}~ This model allows a customer to rent processing power and storage to launch his own virtual machine. It alleviates the costly process of maintaining own data center. One of the important features is that the customers can scale up according to their requirement. It allows their applications to handle high load smoothly. On the other hand, they can save cost when the demand is low. Customers have full control over operating systems, storage, deployed applications, and possibly limited control of selecting networking components (e.g., host firewalls). An example of IaaS is Amazon EC2 \cite{websiteec2}. EC2 provides users with access to virtual machines (VM) running on its servers. Customers can install any operating system and can run any application in that VM. It also gives the customers the facility of saving the VM status by creating an image of the instance. The VM can be restored later by using that image.\smallskip

\noindent\textbf{Other service models.}~Motahari-Nezhad et al. proposed a more specific service model, which is Database as a Service (DaaS) \cite{motahari2009outsourcing}. This is a special type of storage service provided by the cloud service provider. Most of the providers offer the customers to store data in a key-value pair, rather than using traditional relational database. Also data of multiple users can be co-located in a shared physical table. Two of the examples of DaaS are: Amazon SimpleDB \cite{websiteamazonsimpledb} and Google Bigtable \cite{chang2008bigtable}. The query language to store, retrieve, and manipulate the data depends on the provider. There is a monthly fee depending on the incoming and outgoing volume of data and machine utilization.\\

\noindent\textbf{Classification according to deployment model.~}According to the deployment model, cloud computing can be categorized into four categories -- private cloud, public cloud, community cloud, and hybrid cloud \cite{mell2009draft}.\smallskip

\noindent \textbf{Private cloud.~} In private cloud model, the cloud infrastructure is fully operated by the owner organization. It is the internal data center of a business organization. Usually, the infrastructure is located at the organizations' premise. Private cloud can be found in large companies and for research purpose.\smallskip

\noindent \textbf{Community cloud.~} If several organizations with common concerns (e.g., mission, security requirements, policy, and compliance considerations) share cloud infrastructure then this model is referred as community cloud.\smallskip

\noindent \textbf{Public cloud.~} In the public cloud model, the Cloud Service Providers (CSP) owns the cloud infrastructure and they make it available to the general people or a large industry group. All the examples given in the service based cloud categorization are public cloud.\smallskip

\noindent \textbf{Hybrid cloud.~}  As the name suggests, the hybrid cloud infrastructure is a composition of two or more clouds (private, community, or public).(e.g., cloud bursting for load-balancing between clouds). Hybrid Cloud architecture requires both on-premises resources and off-site (remote) server based cloud infrastructure.

\begin{figure*}
\centering
\includegraphics[width=0.98\textwidth]{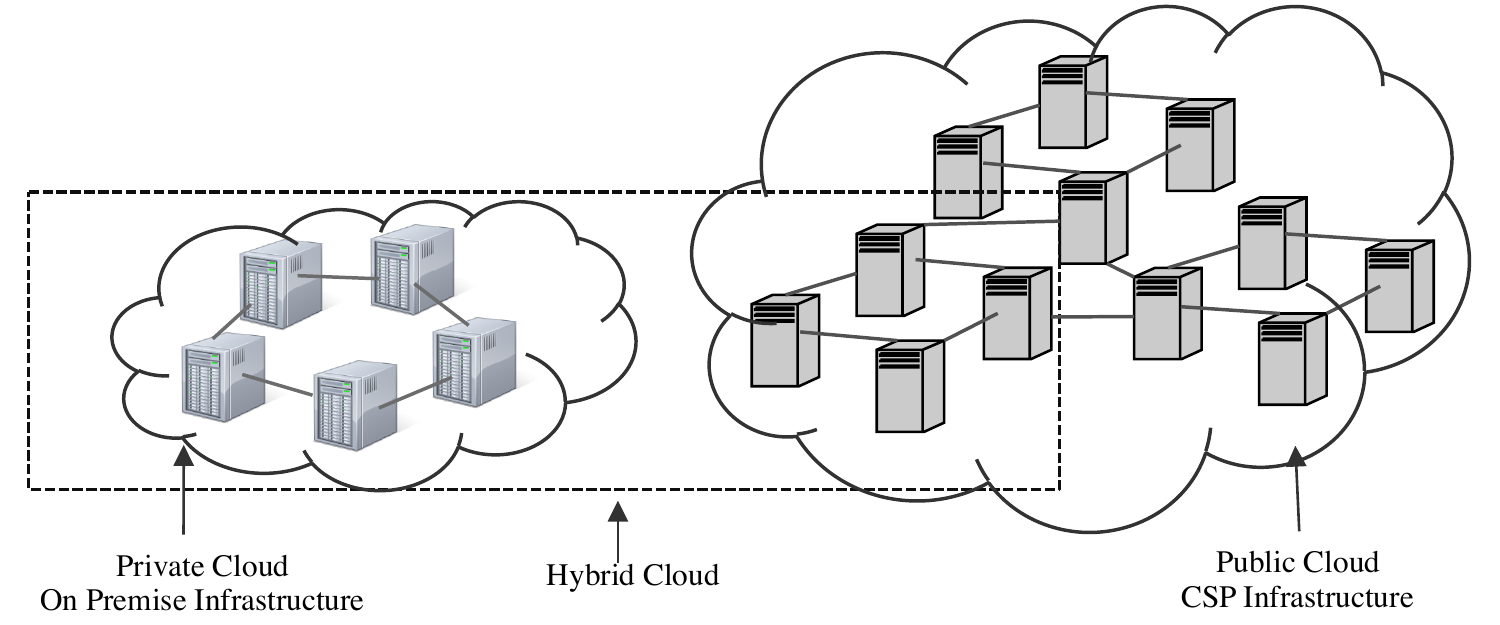}
\caption{Three different cloud deployment models}
\label{figure:clouddeploymenttype}
\end{figure*}

Figure \ref{figure:clouddeploymenttype} shows three different deployment models of cloud computing -- private, public, and hybrid cloud.

\subsection{Computer Forensics}
Computer forensics is the process of preserving, collecting, confirming, identifying, analyzing, recording, and presenting crime scene information. Wolfe defines computer forensics as \emph{``a methodical series of techniques and procedures for gathering evidence, from computing equipment and various storage devices and digital media, that can be presented in a court of law in a coherent and meaningful format"} \cite{wiles2007best}.  According to a definition from NIST \cite{kent2006guide}, computer forensic is \textit{``an applied science to identify a incident, collection, examination, and analysis of evidence data''}. In computer forensics, maintaining the integrity of the information and strict chain of custody for the data is mandatory. Several other researchers define computer forensic as the procedure of examining computer system to determine potential legal evidence \cite{lunn2000computer,robbins2008explanation}. 

From the definitions, we can say that computer forensics is comprised of four main processes:

\begin{itemize}
\item \emph{Identification:} Identification process is comprised of two main steps: identification of an incident and identification of the evidence, which will be required to prove the incident. 
\item \emph{Collection:} In the collection process, an investigator extracts the digital evidence from different types of media e.g., hard disk, cell phone, e-mail, and many more. Additionally,  he needs to preserve the integrity of the evidence. 
\item \emph{Organization:} There are two main steps in organization process: examination and analysis of the digital evidence. In the examination phase, an investigator extracts and inspects the data and their characteristics. In the analysis phase, he interprets and correlates the available data to come to a conclusion, which can prove or disprove civil, administrative, or criminal allegations. 
\item \emph{Presentation:} In this process, an investigator makes an organized report to state his findings about the case. This report should be appropriate enough to present to the jury.
\end{itemize}

\begin{figure*}
\centering
\includegraphics[width=0.98\textwidth]{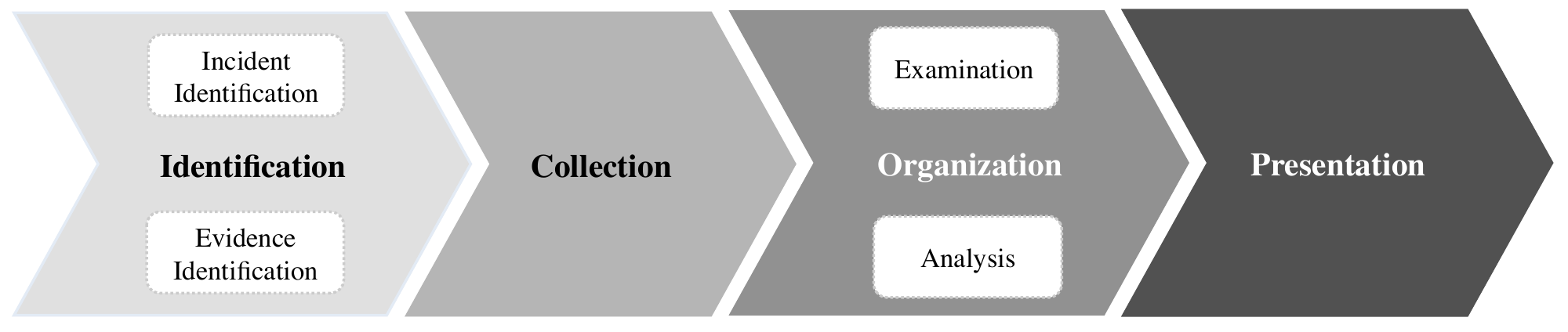}
\caption{Computer Forensics Process Flow}
\label{figure:computerforensic}
\end{figure*}

Figure \ref{figure:computerforensic} illustrates the flow of aforementioned processes in computer forensics.\smallskip

\noindent\textbf{Legal basis.~}Before 2006, there had been no separate US Federal law for computer forensics investigation in civil cases. For criminal cases, investigators still use the 1986 Computer Fraud and Abuse Prevention Act. As computer based crime was increasing rapidly, the Advisory Committee on Civil Rules took initiative to resolve this issue at 2000. Finally at 2006, an amendment to Federal Rules of Civil Procedure (FRCP) was published, which is known as e-discovery amendment \cite{websiteediscoveryammendment}. Some important factors in the FRCP amendment, that are contributing in today's digital forensics are:
\begin{itemize}
	\item FRCP defines the discoverable material and introduces the term \emph{Electronically Stored Information (ESI)}. Under this definition, data stored in hard disk, RAM, or Virtual Machine (VM) logs, all are discoverable material for the forensic investigation.
	\item It introduces data archiving requirements.
	\item It addresses the issue of format in production of ESI. If the responding party objects about the requested format, then it suggests a model for resolving dispute about the form of production.
	\item It provides a Safe Harbor Provision. Under the rule of safe harbor, if someone loses data due to routine faithful operation, then the court may not impose sanction on her for failing to provide ESI. \cite{wiles2007best,websiteediscoveryammendment}.
\end{itemize}

\subsection{Cloud forensics}
We define Cloud forensics as the application of computer forensic principles and procedures in a cloud computing environment. Since cloud computing is based on extensive network access, and as network forensics handles forensic investigation in private and public network, Ruan et al. defined cloud forensics as a subset of network forensics \cite{ruan2011cloud}. They also identified three dimensions in cloud forensics -- technical, organizational, and legal. According to the authors' knowledge,  till now this is only definition of cloud forensics. 

Cloud forensics procedures will vary according to the service and deployment model of cloud computing. For SaaS and PaaS, we have very limited control over process or network monitoring. Whereas, we can gain more control in IaaS and can deploy some forensic friendly logging mechanism. The first three steps of computer forensics will vary for different services and deployment models. For example, the collection procedure of SaaS and IaaS will not be same.  For SaaS, we solely depend on the CSP to get the application log, while in IaaS, we can acquire the Virtual machine instance from the customer and can enter into examination and analysis phase. On the other hand, in the private deployment model, we have physical access to the digital evidence, but we merely can get physical access to the public deployment model. 

\section{Challenges of cloud forensics}
\label{sec:challenge}
In this section, we examine the challenges in cloud forensics, as discussed in the current research literature. We present our analysis by looking into the challenges faced by investigators in each of the stages of computer forensics (as described in Section II-B). Some of the important challenges we address here are: forensic data acquisition, logging, preserving chain of custody, limitation of current forensics tools, crime scene reconstruction, cross border law, and presentation. 

\subsection{Forensic Data Acquisition}
Collection of the digital evidence is the most crucial step of forensic procedure. Any errors that have occurred in the collection phase will propagate to the evidence organization and reporting phase, which will eventually affect the whole investigation process. According to Birk, evidence can be available in three different states in cloud -- at rest, in motion, and in execution \cite{birk2011technical}. Data that occupies the disk space is called data at rest. Data that can be transferred from one state to another state is referred to as data in motion. Sometimes, we have executable data, for example, image snapshot. We can load and run an image snapshot to get the data in rest and data in motion.  In cloud forensics, data collection procedure also varies depending on the service and deployment model of clouds.

Some of the factors that make the data acquisition process in cloud forensic harder than traditional computer forensics are discussed below.\smallskip

\begin{figure*}[tbp]
\centering
\includegraphics[width=1.0\textwidth]{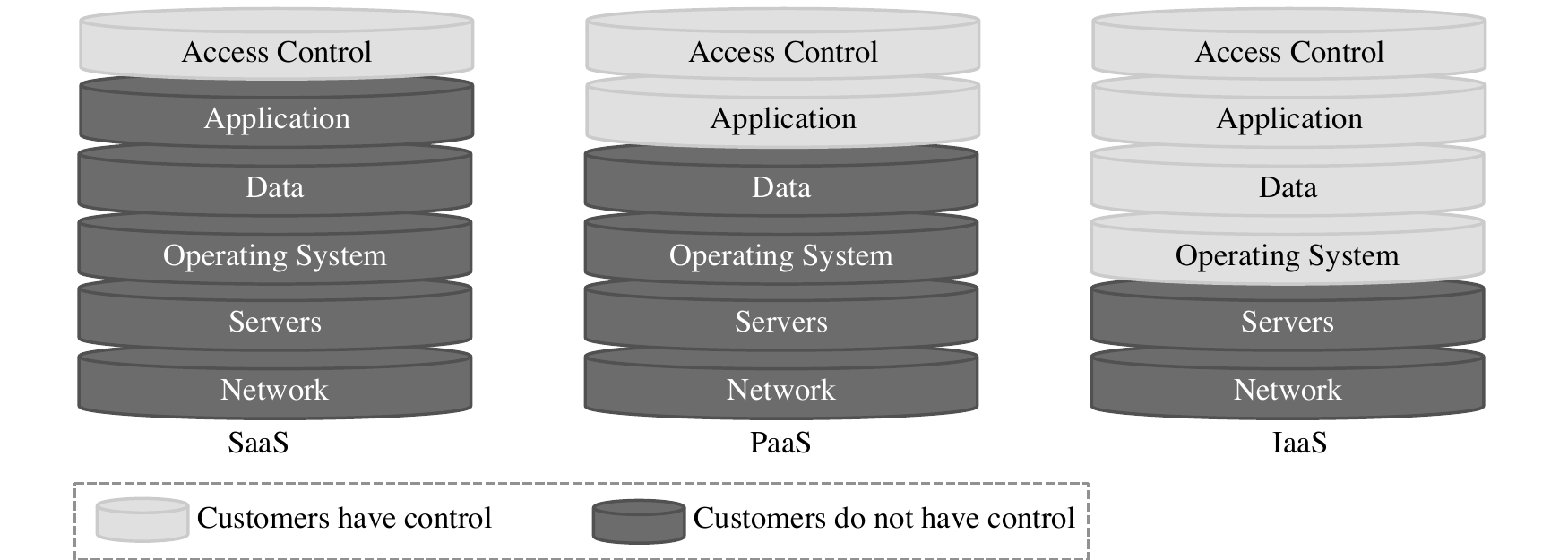}
\caption{Customers' control over different layers in different service model}
\label{figure:servicecontrol}
\end{figure*}

\noindent \textbf{Physical Inaccessibility.~} Physical inaccessibility of digital evidence makes the evidence collection procedure harder in cloud forensics. The established digital forensic procedures and tools assume that we have physical access to the computers. However, in cloud forensics, the situation is different. Sometimes, we do not even know where the data is located as it is distributed among many hosts in multiple data centers. A number of researchers address this issue in their work \cite{birk2011technical,dykstraunderstanding,guoforensicforensic,wolthusen2009overcast,reilly2011cloud,sluskyforensic}. \smallskip

\noindent \textbf{Less Control in Clouds and Dependence on the CSP.~} In traditional computer forensics, investigators have full control over the evidence (e.g., a hard drive confiscated by police). In a cloud, unfortunately, the control over data varies in different service models. Figure \ref{figure:servicecontrol} shows the limited amount of control that customers have in different layers for the three service models -- IaaS, PaaS, and SaaS. For this reason, we mostly depend on the CSP to collect the digital evidence from cloud computing environment. This is a serious bottleneck in the collection phase. 

In IaaS, users have more control than SaaS or PaaS. The lower level of control has made the data collection in SaaS and PaaS more challenging than in IaaS. Sometimes, it is even impossible. If we manage to get the image of an IaaS instance, it will make our life easy to investigate the system. For SaaS and PaaS, we need to depend on the CSP. We can only get a high level of logging information from this two service models. As customers have control over the application deployed in PaaS, they can keep log of different actions to facilitate the investigation procedure. On the contrary in SaaS, customers basically have no control to log the actions. 

Dykstra et al. presented the difficulty of data acquisition by using a hypothetical case study of child pornography  \cite{dykstraunderstanding}. To investigate this case, the forensics examiner needs a bit-for-bit duplication of the data to prove the existence of contraband images and video, but in a cloud, he cannot collect data by himself. At first, he needs to issue a search warrant to the cloud provider. However, there are some problems with the search warrant in respect of cloud environment. For example, warrant must specify a location, but in cloud the data may not be located at a precise location or a particular storage server. Furthermore, the data can not be seized by confiscating the storage server in a cloud, as the same disk can contain data from many unrelated users. To identify the criminal, we need to know whether the virtual machine has a static IP. Almost in all aspects, it depends on the transparency and cooperation of the cloud provider. \smallskip

\noindent \textbf{Volatile Data.~} Volatile data cannot sustain without power. When we turn off a Virtual Machine (VM), all the data will be lost if we do not have the image of the instance. This issue is highlighted in several research works \cite{birk2011technical,guoforensicforensic,wolthusen2009overcast,reilly2011cloud,taylor2010digital}. Though IaaS has some advantages over SaaS and PaaS, volatile storage can be a problem in IaaS model if data is not always synchronized in persistent storage, such as, Amazon S3 or EBS. If we restart or turn off a VM instance in IaaS (e.g., in Amazon EC2), we will lose all the data. Registry entries or temporary internet files, that reside or be stored within the virtual environment will be lost when the user exits the system. Though with extra payment customers can get persistent storage, this is not common for small or medium scale business organizations. Moreover, a malicious user can exploit this vulnerability. After doing some malicious activity (e.g., launch DoS attack, send spam mail), an adversary can power off her virtual machine instance, which will lead to a complete loss of the volatile data and make the forensic investigation almost impossible.  Birk also mentioned a serious problem regarding the volatile nature of evidence in cloud. The problem states that some owner of a cloud instance can fraudulently claim that her instance was compromised by someone else and had launched a malicious activity. Later, it will be difficult to prove her claim as false by a forensic investigation \cite{birk2011technicalIssues}.\smallskip

\noindent \textbf{Trust Issue.~} Dependence on the third party also poses trust issue in investigation procedure. In the child pornography case study, Dykstra et al. highlighted the trust issue in collecting evidence \cite{dykstraunderstanding}. After issuing a search warrant, the examiner needs a technician of the cloud provider to collect data. However, the employee of the cloud provider who collects data is most likely not a licensed forensics investigator and it is not possible to guarantee his integrity in a court of law \cite{sluskyforensic}. The date and timestamps of the data are also questionable if it comes from multiple systems. Dykstra et al. experimented with collecting evidence from cloud environment. One of the shortcomings they found is that it is not possible to verify the integrity of the forensic disk image in Amazon's EC2 cloud because Amazon does not provide checksums of volumes, as they exist in EC2. \smallskip
 
\noindent \textbf{Large Bandwidth:} In Section \ref{sec:introduction}, we have seen that the amount of digital evidence is increasing rapidly. Guo et al. pointed out the requirement of large bandwidth issue for time critical investigation \cite{guoforensicforensic}. The on-demand characteristic of cloud computing will have vital role in increasing the digital evidence in near future. In traditional forensic investigation, we collect the evidence from the suspect's computer hard disk. Conversely, in cloud, we do not have physical access  to the data. One way of getting data from cloud VM is downloading the VM instance's image. The size of this image will increase with the increase of data in the VM instance. We will require adequate bandwidth and incur expense to download this large image. \smallskip

\noindent \textbf{Multi-tenancy.~} In cloud computing, multiple VM can share the same physical infrastructure, i.e., data for multiple customers may be co-located. This nature of clouds is different from the traditional single owner computer system. In any adversarial case, when we acquire evidence two issues can arise. First, we need to prove that data were not co-mingled with other users' data \cite{dykstraunderstanding,guoforensicforensic}. And secondly, we need to preserve the privacy of other tenants while performing an investigation \cite{sluskyforensic}. Both of these issues make acquiring digital evidence  more challenging. The multi-tenancy characteristic also brings the side-channel attacks \cite{ristenpart2009hey} that are difficult to investigate.

\subsection{Logging} 
Analyzing logs from different processes plays a vital role in digital forensic investigation. Process logs, network logs, and application logs are really useful to identify a malicious user. However, gathering this crucial information in cloud environment is not as simple as it is in privately owned computer system, sometimes even impossible. Cloud forensic researchers have already identified a number of challenges in cloud based log analysis and forensics \cite{guoforensicforensic,sluskyforensic,marty2011cloud}. We briefly discuss these challenges below.\\\smallskip
   
\noindent \textbf{Decentralization.~} In cloud infrastructure, log information is not located at any single centralized log server; rather logs are decentralized among several servers. Multiple users' log information may be co-located or spread across multiple servers. \smallskip

\noindent \textbf{Volatility of Logs.~} Some of the logs in cloud environment are volatile, especially  in case of VM. All the logs will be unavailable if the user power off the VM instance. Therefore, logs will be available only for certain period of time.\smallskip

\noindent \textbf{Multiple Tiers and Layers.~} There are several layers and tiers in cloud architecture. Logs are generated in each tier. For example, application, network, operating system, and database -- all of these layers produce valuable logs for forensic investigation. Collecting logs from these multiple layers is challenging for the investigators.  \smallskip

\noindent \textbf{Accessibility of Logs.~} The logs generated in different layers are need to accessible to different stakeholders of the system, e.g., system administrator, forensic investigator, and developer. System administrators need relevant log to troubleshoot the system. Developers need the required log to fix the bug of the application. Forensic investigators need logs that can help in their investigation. Hence, there should be some access control mechanism, so that everybody will get what they need exactly -- nothing more, nothing less and obviously, in a secure way. \smallskip

\noindent \textbf{Dependence on the CSP.~} Currently, to acquire the logs, we extensively depend on the CSPs. The availability of the logs varies depending on the service model. In SaaS, customers do no get any log of their system, unless the CSP provides the logs. In PaaS, it is only possible to get the application log from the customers. To get the network log, database log, or operating system log we need to depend on the CSP. For example, Amazon does not provide load balancer log to the customers \cite{websiteamazonwebservice}. In a recent research work, Marty mentioned that he was unable to get MySql log data from Amazon's Relational Database Service \cite{marty2011cloud}. In IaaS, customers do not have the network or process log. \smallskip

\noindent \textbf{Absence of Critical Information in Logs.~} There is no standard format of logs. Logs are available in heterogeneous formats -- from different layers and from different service providers. Moreover, not all the logs provide crucial information for forensic purpose, e.g., who, when, where, and why some incident was executed.

\subsection{Chain of Custody}
Chain of custody is defined as a verifiable provenance or log of the location and possession history of evidence from the point of collection at the crime scene to the point of presentation in a court of law. It is one of the most vital issues in traditional digital forensic investigation. Chain of custody should clearly depicts how the evidence was collected, analyzed, and preserved in order to be presented as admissible evidence in court \cite{vacca2005computer}. In traditional forensic procedure, it starts with gaining the physical control of the evidence, e.g., computer, and hard disk. However, in cloud forensics, this step is not possible. In a cloud, investigator can acquire the available data from any workstation connected with the internet. Due to the multi jurisdictional laws, procedures, and proprietary technology in cloud environment,  maintaining chain of custody will be a challenge \cite{taylor2010digital,grisposcalm}. In a hypothetical case study of compromised cloud based website, Dykstra et al. pointed that as multiple people may have access to the evidence and we need to depend on the CSP to acquire the evidence, the chain of custody preservation throughout the investigation process is  questionable \cite{dykstraunderstanding}. According to Birk et al. the chain of custody will be a problem in cloud forensic as the trustworthiness of hypervisor is also questionable \cite{birk2011technical}.

\subsection{Limitations of Current of Forensic Tools}
Due to the distributed and elastic characteristic of cloud computing, the available forensic tools cannot cope up with this environment. Some researchers highlighted the limitations of current forensic tools in their work \cite{ruan2011cloud,reilly2011cloud,grisposcalm}. Tools and procedures are yet to be developed for investigations in virtualized environment, especially on hypervisor level. Ruan et al. expressed the need of forensic-aware tools for the CSP and the clients to collect forensic data \cite{ruan2011cloud}. 

\subsection{Crime Scene Reconstruction}
To investigate a malicious activity, sometimes the investigators need to reconstruct the crime scene. It helps them to understand how an adversary launched the attack. However, in cloud environment, that could be a problem \cite{reilly2011cloud}. If an adversary shut down her virtual instance after a malicious activity or undeploy his malicious website, then reconstruction of the crime scene will be impossible.

\subsection{Cross Border Law}
Multi-jurisdictional or cross border law is intensifying the challenge of cloud forensics. Data centers of the service providers are distributed worldwide. However, the privacy preservation or information sharing laws are not in harmonic throughout the world, even it may not be same in different states of a country. Cross border legislation and cross border red tape issues came in several cloud forensic research works \cite{ruan2011cloud,zafarullah2011digital,biggs2009cloud} which make the evidence collection process challenging. In particular, such a process should not violate the laws of a particular jurisdiction. Furthermore, the guideline of admissible evidence, or the guideline for preserving chain of custody can vary among different regions. It may happen that the attacker is accessing the cloud computing service from one jurisdiction, whereas the data she is accessing reside in different jurisdiction. Differences in laws between these two locations can affect the whole investigation procedure, from evidence collection, presenting proofs to capture the attacker. Moreover, for multi-tenancy case, we need to preserve the privacy of the tenants when we collect data of other tenant, sharing the same resources. However, the privacy and privilege rights may vary among different countries or states.  

\subsection{Presentation}
The final step of digital forensic investigation is presentation, where an investigator accumulates his findings and presents to the court as the evidence of a case. Challenges also lie in this step of cloud forensics. Proving the evidence in front of the jury for traditional computer forensics is relatively easy compared to the complex structure of cloud computing. Jury members possibly have basic knowledge of personal computers or at most privately owned local storage. But the technicalities of a cloud data center, running thousands of VM, accessed simultaneously by hundreds of users is far too complex for them to understand \cite{reilly2011cloud}.

\begin{table}[tbp]
\begin{center}
	\renewcommand{\arraystretch}{1}
	\begin{tabular}{|p{0.17\textwidth} | p{0.025\textwidth} | p{0.025\textwidth} | p{0.025\textwidth} | p{0.11\textwidth} |}
	\hline
	
	\cellcolor[gray]{0} \textcolor{white}{Challenges of Cloud Forensics} & \multicolumn{3}{|c|}{\cellcolor[gray]{0}\textcolor{white} {Exists in}} & \cellcolor[gray]{0}\textcolor{white}{Work}\\ \hline 
	 & \cellcolor[gray]{0} \textcolor{white}{IaaS} & \cellcolor[gray]{0} \textcolor{white}{PaaS} & \cellcolor[gray]{0} \textcolor{white}{SaaS} & \\ \hline
	
	Physical inaccessibility & \ding{51} & \ding{51} & \ding{51} & \cite{birk2011technical,dykstraunderstanding,guoforensicforensic,wolthusen2009overcast,sluskyforensic,reilly2011cloud}.   \\ \hline
	
	Dependence on CSP & \ding{51} & \ding{51} & \ding{51} & \cite{dykstraunderstanding} \\ \hline
	Volatile Data & \ding{51} & \ding{53} & \ding{53} & \cite{birk2011technical,guoforensicforensic,wolthusen2009overcast,taylor2010digital,reilly2011cloud} \\ \hline
	Trust Issue & \ding{51} & \ding{51} & \ding{51} & \cite{dykstraunderstanding,sluskyforensic} \\ \hline
	Large bandwidth & \ding{51} & \ding{53} & \ding{53} & \cite{guoforensicforensic} \\ \hline
	Multi-tenancy & \ding{51} & \ding{51} & \ding{53} & \cite{dykstraunderstanding,guoforensicforensic,sluskyforensic} \\ \hline
	Decentralization of Logs & \ding{51} & \ding{51} & \ding{51} & \cite{marty2011cloud,guoforensicforensic} \\ \hline
	Volatility of logs & \ding{51} & \ding{53} & \ding{53} & \cite{marty2011cloud,sluskyforensic} \\ \hline
	Logs in multiple tiers and layers & \ding{51} & \ding{51} & \ding{51} & \cite{marty2011cloud} \\ \hline
	Accessibility of logs & \ding{51} & \ding{51} & \ding{51} & \cite{marty2011cloud} \\ \hline
	Depending on CSP for logs & \ding{51} & \ding{51} & \ding{51} & \cite{guoforensicforensic} \\ \hline
	Absence of critical information in logs & \ding{51} & \ding{51} & \ding{51} & \cite{marty2011cloud} \\ \hline
	Chain of Custody & \ding{51} & \ding{51} & \ding{51} & \cite{taylor2010digital,grisposcalm,dykstraunderstanding,birk2011technical} \\ \hline
	Problem of current forensic tools & \ding{51} & \ding{51} & \ding{51} & \cite{ruan2011cloud,reilly2011cloud,grisposcalm} \\ \hline
	Crime scene reconstruction & \ding{51} & \ding{51} & \ding{53} & \cite{reilly2011cloud} \\ \hline
	Cross border law & \ding{51} & \ding{51} & \ding{51} & \cite{zafarullah2011digital,biggs2009cloud,taylor2010digital,ruan2011cloud} \\ \hline
	
	Presentation & \ding{51} & \ding{51} & \ding{51} & \cite{reilly2011cloud} \\ \hline
	Compliance issue & \ding{51} & \ding{51} & \ding{53} & \cite{reilly2011cloud} \\ \hline
	
	\end{tabular}
\end{center}
\caption{Summary of Challenges in Cloud Forensics}
\label{table:challenge}
\end{table}
\subsection{Trustworthy data retention}
Large business organization and medicals cannot move to cloud because of some compliance issues. Trustworthy data retention is one of the mandatory compliance issues that has direct impact on digital forensics. Hasan et al. state that trustworthy data retention should provide the long-term retention and disposal of organizational record to prevent unwanted deletion, editing, or modification of data during the retention period. It should also prevent recreation of record once it has been removed \cite{gertz2008handbook}. While there are still some open problems to ensure the secure data retention at storage level, the cloud computing model imposes some new challenges. Popovic et al. mentioned some issues about retention and destruction of record in cloud computing. For example, who enforces the retention policy in the cloud, and how are exceptions, such as, litigation holds managed? Moreover, how can the CSPs assure us that they do not retain data after destruction of it \cite{popovic2010cloud}? There are several laws in different countries, which mandate the trustworthy data retention. Just in United States, there are 10,000 laws at the federal and state levels that force the organizations to manage records securely \cite{gertz2008handbook}. Some of the laws and regulations are stated below:
\begin{itemize}
	\item \emph{Sarbanes-Oxley Act:} This act mandates public companies to provide disclosure and accountability of their financial reporting, subject to independent audits \cite{websitesarbanesox}.
	\item \emph{The Health Insurance Portability and Accountability Act (HIPAA):} This act requires privacy and confidentiality of patient medical record \cite{websitehippa}.
	\item \emph{The Securities and Exchange Commission (SEC) rule 17a-4:} According to this law, traders, brokers, and financial companies need to maintain their business records, transactions, and communications for a number of years \cite{websitesec}.
	\item \emph{Federal Information Security Management Act:} This law regulates information systems used by the Federal government and affiliated parties \cite{websiteuscongress}.
	\item \emph{The Gramm-Leach-Bliley Act:}According to this law, financial institutions must have a policy to protect information from any predictable threats in integrity and data security \cite{grammleachblileyact}.
	\item \emph{European Commission data protection legislation:} In 2012, European Commission proposes major reformation of the 1995 data protection legislation to strengthen the privacy and confidentiality of individuals' data \cite{websiteeuparliament}.\\	
\end{itemize}

All of the above laws mandate trustworthy data retention. Compliance with all of these laws is challenging in cloud computing environment. For example, the SOX act mandates that the financial information must be resided in an auditable storage, which the CSPs do not provide. Business organization cannot move their financial information to cloud infrastructure as it does not comply with the act. As cloud infrastructure does not comply with HIPAA's forensic investigation requirement, hospitals cannot move the patients' medical information to cloud storage. As business and healthcare organizations  are the two most data consuming sectors, cloud computing cannot achieve the ultimate success without including these two sectors. These sectors are spending extensively to make their own regulatory-compliant infrastructure. A regulatory-compliant cloud can save this huge investment. Hence, we need to solve the audit compliance issue to bring more customers in cloud world.\\\smallskip

	Table~\ref{table:challenge} gives an overview of the challenges in three service models of cloud computing for publicly deployed cloud.\\

\section{Current Solutions}
\label{sec:solution}
In this section, we discuss some existing proposed solutions, which can mitigate some of the challenges of cloud forensics.
\subsection{Trust Model}
In Section \ref{sec:challenge}, we have already seen that for forensic data acquisition, we need to depend on the CSP heavily. This inevitably affects trust and evidence integrity. Dykstra et al. proposed a trust model with six layers: Guest application / data, Guest OS, Virtualization, Host OS, Physical hardware, and Network. The further down the stack is, the less cumulative trust is required. For example, in Guest OS layer, we require trust in Guest OS, hypervisor, host OS, hardware, and network layer. While, in network layer, we require trust in only the network. Examiners can examine evidence from different layer to ensure the consistency of the digital evidence. For forensic examination, we need to choose the appropriate layer, which depends on the data available in the layer and trust in the available data \cite{dykstraacquiring}. Wolthusen suggested an interactive evidence presentation and visualization mechanism to overcome the trust issue \cite{wolthusen2009overcast}. Ko et al. proposed TrustCloud -- a trust preserving framework for cloud \cite{ko2011trustcloud}.

\subsection{Integrity Preservation}
Integrity preservation of the digital evidence is a crucial step in cloud forensic investigation. Without integrity preservation, the validity of the evidence will be questionable and the jury can object about it. Generating a digital signature on the collected evidence and then checking the signature later is one way to validate the integrity. As data is distributed among multiple servers, this procedure is not simple, rather quite complicated. However, cloud researchers proposed some mechanisms to generate and check the signature of distributed cloud data.

Hegarty et al. also proposed a distributed  signature detection  framework that will facilitate the forensic investigation in cloud environment \cite{hegartyforensic}. Traditional techniques of signature detection for digital forensic are not efficient and appropriate due to the distributed nature of cloud computing. Current model of file storage is comprised of two components -- Meta data Servers (MDS) and Object Storage Devices (OSD). The hash value of each file is stored in the MDS as an e-tag and integrity is checked each time after uploading / downloading a file. In the proposed framework, the first step is to send a list of target buckets to the  Forensic  Cluster  Controller (FCC), along  with a file containing the target MD5 hash values. The FCC then initializes and queries to Analysis Nodes (AN) for getting the number of files contained in targeted bucket. Upon receiving the round one signature file from FCC, each AN retrieves the e-tags of the bucket. The signatures in the round one signature file are compared with the signatures generated from the e-tags by the AN. After getting feedback from all ANs, FCC terminates the ANs. They tested their framework by two ways -- using Amazon S3 and by emulating a cloud platform. They achieved zero false positive and false negative rate and found significant improvement in terms of data required at AN.

\subsection{Logging}
Log information is vital for forensic investigations. A lot of researchers have explored logging in the context of a cloud. Marty proposed a log management solution, which can solve several challenges of logging as discussed in Section~\ref{sec:challenge} \cite{marty2011cloud}. In the first step of the logging solution, logging must be enabled on all infrastructure components to collect logs. The next step is for establishing a synchronized, reliable, bandwidth efficient, and encrypted transport layer to transfer log from the source to a central log collector. The final step deals with ensuring the presence of the desired information in the logs. The proposed guideline tells us to focus on three things: when to log, what to log, and how to log. The answer to when to log depends on the use cases, such that business relevant logging, operations based logging, security (forensics) related logging, and regulatory and standards mandates. At minimum, he suggested to log the timestamps record, application, user, session ID, severity, reason, and categorization, so that we can get the answer of what, when, who, and why (4 W). He also recommended syntax for logging, which was represented as a key-value pair and used three fields to establish a categorization schema -- object, action, and status. He also implemented an application logging infrastructure at a SaaS company, where he built a logging library that can be used in Django. This library can export logging calls for each severity level, such as, debug, info, error, and others. For logging in javascript layer, he built an AJAX library to store the logs in server side. Then he tuned the apache configuration to get the logs in desired format and to get the logs from load balancer. For logging the back-end operations, he used Log4j as the backend was built in java. While there are several advantages to this approach, this work does not provide any solution about logging network usage, file metadata, process usage, and many other important evidence, which are important for forensic investigation in IaaS and PaaS models. 
	
To facilitate logging in clouds, Zafarullah et al. proposed logging provided by OS and the security logs \cite{zafarullah2011digital}. In order to investigate  digital forensics in cloud, they set up cloud computing environment by using Eucalyptus. Using Snort, Syslog, and Log Analyzer (e.g., Sawmill), they were able to monitor the Eucalyptus behavior and log all internal and external interaction of Eucalyptus components. For their experiment, they launched a DDoS attack from two virtual machine and analyzed bandwidth usage log and processor usage log to detect the DDoS attack. From the logs in \emph{/var/eucalyptus/jetty-request-05-09-xx} file on Cloud Controller (CC) machine, it is possible to identify the attacking machine IP, browser type, and content requested. From these logs, it is also possible to determine the total number of VMs, controlled by single Eucalyptus user and VMs communication patterns. Their experiment shows that if the CSPs come forward to provide better logging mechanism, cloud forensics will be benefited greatly.  

	To get necessary logs from all the three service models, Bark et al. proposed that CSP could provide network, process and access logs to customer by read only API \cite{birk2011technicalIssues}. By using these APIs, customer can provide valuable information to investigator. In PaaS, customers have full control on their application and can log variety of access information in a configurable way. Hence for PaaS, they proposed a central log server, where customer can store the log information. In order to protect log data from possible eavesdropping and altering action, customers can encrypt and sign the log data before sending it to the central server.
	
\subsection{Cloud Management Plane}
	Data acquisition from cloud infrastructure is a challenging step in cloud forensics. CSPs can play a vital role in this step by providing a web based management console like AWS management console. Dykstra et al. recommended a cloud management plane for use in IaaS model \cite{dykstraacquiring}. From the console panel, customers as well as investigators can collect VM image, network, process, database logs, and other digital evidence, which cannot be collected in other ways. Only problem with this solution is that, it requires an extra level of trust -- trust in the management plane. In traditional evidence collection procedure, where we have physical access to the system, this level of trust is not required. 

\subsection{Solution of Legal Issues}
	Legal issue is a great obstacle in cloud forensics. Cross border legislations often hinder the forensic procedures. At present, there is a massive gap in the existing Service Level Agreement (SLA), which neither defines the responsibility of CSPs at the time of some malicious incident, nor their role in forensic investigation. Researches gave emphasis on sound and robust SLA between cloud service providers and customers  \cite{dykstraunderstanding,biggs2009cloud,birk2011technicalIssues}. To resolve the transparency issues, the CSP should build a long-term trust relationship with customers. A robust SLA should state how the providers deal with the cyber crimes, i.e., how and to which extent they help in forensic investigation procedure. In this context, another question can come -- how we can be sure of the robustness of a SLA. To ensure the quality of SLA, we can take help from a trusted third party \cite{birk2011technicalIssues}. To overcome the cross border legislation challenges, Biggs proposed an international unity for introducing an international legislation for cloud forensics investigation \cite{biggs2009cloud}. By implementing a global law throughout the world, we can make the investigation procedure smooth enough to complete in a time limit. 

\subsection{Virtual Machine Introspection}
	Virtual Machine Introspection (VMI) is the process of externally monitoring the runtime state of VM from either the Virtual Machine Monitor (VMM), or from some virtual machine other than the one being examined. By runtime state, we are referring to processor registers, memory, disk, network, and other hardware-level events. Through this process, we can execute a live forensic analysis of the system, while keeping the target system unchanged \cite{hay2008forensics}. In this work, Hay et al. showed that if a VM instance is compromised by installing some rootkit to hide the malicious events, it is still possible to identify those malicious events by performing VMI. They used an open source VMI library, Xen (VIX) suite to perform their experiment. However, this tool is no longer maintained under this name, it is now known as LibVMI \cite{websitelibvmi}.  

\subsection{Continuous Synchronization}
In order to provide the on demand computational and storage service, CSPs do not provide persistent storage to VM instance. If we turn off the power or reboot the VM, we will eventually lose all the data reside in the VM. To overcome the problem of volatile data, Birk et al. mentioned about the possibility of continuous synchronization of the volatile data with a persistent storage \cite{birk2011technicalIssues}. However, they did not provide any guideline about the procedure.
	There can be two possible ways of continuous synchronization. 
	\begin{itemize}
		\item CSPs can provide a continuous synchronization API to customers. Using this API, customers can preserve the synchronized data to any cloud storage e.g., Amazon S3, or to their local storage. Implementing this mechanism will be helpful to get the evidence from a compromised VM, even though the adversary shut down the VM after launching any malicious activity. 
		\item However, if the adversary is the owner of a VM, the above-mentioned mechanism will not work. Trivially, she will not be interested to synchronize her malicious VM. To overcome this issue, CSPs by themselves can integrate the synchronization mechanism with every VM and preserve the data within their infrastructure.   
	\end{itemize}

\begin{table}[tbp]
\begin{center}
	\renewcommand{\arraystretch}{1}
	\begin{tabular}{|p{0.18\textwidth} | p{0.03\textwidth} | p{0.03\textwidth} | p{0.03\textwidth} | p{0.09\textwidth} |}
	\hline
	\cellcolor[gray]{0}\textcolor{white}{ Proposed Solution} & \multicolumn{3}{|c|}{\cellcolor[gray]{0}\textcolor{white}{ Suitable for}} & \cellcolor[gray]{0}\textcolor{white}{Work}\\ \hline 
	& \cellcolor[gray]{0} \textcolor{white}{IaaS} & \cellcolor[gray]{0} \textcolor{white}{PaaS} & \cellcolor[gray]{0} \textcolor{white}{SaaS} & \\ \hline
	
	Trust Model & \ding{51} & \ding{51} & \ding{53} & \cite{dykstraacquiring,wolthusen2009overcast,ko2011trustcloud} \\ \hline
	Distributed  signature detection framework & \ding{53} & \ding{53} & \ding{51} & \cite{hegartyforensic} \\ \hline
	Log management solution & \ding{53} & \ding{51} & \ding{53} & \cite{marty2011cloud} \\ \hline
	OS and the security logs & \ding{51} & \ding{53} & \ding{53} & \cite{zafarullah2011digital} \\ \hline
	API provide by CSP for logs & \ding{51} & \ding{51} & \ding{51} & \cite{birk2011technicalIssues} \\ \hline
	Cloud management plane & \ding{51} & \ding{51} & \ding{51} & \cite{dykstraacquiring} \\ \hline
	Robust SLA & \ding{51} & \ding{51} & \ding{51} & \cite{dykstraunderstanding,biggs2009cloud,birk2011technicalIssues} \\ \hline
	Trusted Third Party & \ding{51} & \ding{51} & \ding{51} & \cite{birk2011technicalIssues} \\ \hline
	Global unity & \ding{51} & \ding{51} & \ding{51} & \cite{biggs2009cloud} \\ \hline
	Virtual Machine Introspection & \ding{51} & \ding{53} & \ding{53} & \cite{hay2008forensics} \\ \hline
	Continuous synchronization & \ding{51} & \ding{51} & \ding{53} & \cite{birk2011technicalIssues} \\ \hline
	Trusted Platform Module (TPM) & \ding{51} & \ding{51} & \ding{53} & \cite{birk2011technicalIssues,dykstraacquiring} \\ \hline
	Isolating a Cloud Instance & \ding{51} & \ding{53} & \ding{53} & \cite{Delport2011cloud} \\ \hline
	Data Provenance in Cloud & \ding{51} & \ding{51} & \ding{51} & \cite{birk2011technicalIssues,dykstraunderstanding} \\ \hline
	\end{tabular}
\end{center}
\caption{Summary of Solutions in Cloud Forensics}
\label{table:solution}
\end{table}
\begin{table}[tbp]
\begin{center}
	\renewcommand{\arraystretch}{1}
	\begin{tabular}{|p{0.24\textwidth} | p{0.20\textwidth} |}
	\hline
	\cellcolor[gray]{0}\textcolor{white}{Challenges} & \cellcolor[gray]{0} \textcolor{white}{Proposed Solution}\\ \hline
	Trust issue for depending on CSP & Trust Model\\ \hline
	Preserving integrity & Distributed signature detection framework\\ \hline
	Decentralization of logs, logs in multiple tiers and layers, absence of critical information in logs, Volatility of logs & Log management solution\\ \hline
	Depending on CSP for logs & API provide by CSP for logs\\ \hline
	Dependability on CSP for data acquisition & Cloud management plane\\ \hline
	Compliance issue, dependability on CSP & Robust SLA\\ \hline
	Compliance issue, Developing a robust SLA & Trusted third party\\ \hline
	Cross border law & Global unity\\ \hline
	Live forensics issue & Virtual machine introspection\\ \hline
	Volatile Data & Continuous synchronization\\ \hline
	Trust issues of cloud computing & Trusted platform module (TPM)\\ \hline
	Multi-tenancy issue & Isolating a cloud instance\\ \hline
	Chain of custody & Data provenance in cloud\\ \hline
	\end{tabular}
\end{center}
\caption{Analysis of Challenges and Proposed Solutions}
\label{table:challengesolution}
\end{table}
\subsection{Trusted Platform Module (TPM)}
	To preserve the integrity and confidentiality of the data, several researchers proposed TPM as the solution \cite{birk2011technicalIssues,dykstraacquiring}. TPM for cloud computing was proposed by several researchers previously for ensuring trust in cloud computing \cite{krautheim2010introducing,santos2009towards}. By using TPM, we can get machine authentication, hardware encryption, signing, secure key storage, and attestation. It can provide the integrity of the running virtual instance, trusted log files, and trusted deletion of data to customers. However, Dykstra et al. mentioned that TPM is not totally secure and it is possible to modify a running process without being detected by TPM. Moreover, at present, CSPs have heterogeneous hardware and few of them have TPM. Hence, CSPs cannot ensure a homogeneous hardware environment with TPM in near future. 

\subsection{Isolating a Cloud Instance}
	A cloud instance must be isolated if any incident take place on that instance. Isolation is necessary because it helps to protect evidence from contamination. However, as multiple instances can be located in one node, this task becomes challenging. Delport et al. presented some possible techniques of cloud isolation \cite{Delport2011cloud}. Moving a suspicious instance from one node to another node may result in possible loss of evidence. To protect evidence, we can move other instances reside in the same node. The first technique that is proposed is instance relocation. To move an instance, data on the secondary storage, content of the virtual memory, (e.g., swap memory), and the running processes must be moved. Relocation can be done in two ways -- manual and automatic. In the manual mode, the administrator has all the power to move the instance. In automatic mode, the CSP move the instance from one node to another. While moving, the challenge is to ensure confidentiality, integrity,  and availability of other users' data. The second technique is server farming, which can be used to re-routing request between user and node. The third technique is failover, where there is at least one server that is replicating another. There are three ways of failover -- Client-based failover, DNS-based failover and IP-address takeover. Address relocation is another technique, which is actually a special case of DNS-based failover. When it is detected that the main computer has failed, the traffic is rerouted to the backup server. However, this procedure depends on the success of replication. We can also isolate an instance by placing it in a sandbox. One approach of creating a sandbox is installing a sandboxing application in cloud operating system. Another approach is creating a virtual box around an instance and observe all the communication channel. The third technique is placing a Man in the Middle (MITM) between cloud instance and hardware. In that way, we can get log information from CPU, RAM, hard drive, and network. To get benefit from this mechanism, the CSP should embrace this technique for implementation in its cloud.

\subsection{Data Provenance in Cloud}
	Provenance provides the history of an object. By implementing secure provenance, we can get some important forensic information, such as, who owns the data at a given time, who accesses the data, and when. Some researchers have applied the principles of provenance to cloud forensics \cite{dykstraunderstanding,birk2011technicalIssues}. Secure provenance can ensure the chain of custody in cloud forensics as it can provide the chronological access history of evidence, how it was analyzed, and preserved. There have been several projects for secure provenance in cloud computing \cite{muniswamy2010cloud,muniswamy2010provenance}, but no CSP has practically implemented any of the mechanisms yet.\\\smallskip
	
Table~\ref{table:solution} gives an overview of the solution in three different service model of cloud computing. Table~\ref{table:challengesolution} provides an analysis of challenges and solution, i.e. which solution is applicable to overcome which challenge.

\label{sec:others}
\section{Evaluation of current forensic tools in cloud}
\label{sec:evaluation}
There are some popular and proven digital forensics tools used by forensic investigators, e.g., Encase, Accessdata FTK, and others. Though the data acquisition procedure is different in a cloud environment compared to traditional computer forensics, these tools can be used for data acquisition from cloud environment. So far, there has been a single work that evaluates the capability of some available forensic tools in cloud environment \cite{dykstraacquiring}. To evaluate the capability of forensic tool, Dykstra et al. mostly focus on data acquisition step. They chose Amazon EC2 for their experiment, and used EnCase and Accessdata FTK to remotely acquire forensic evidence. They conducted three experiments to collect data from three different layers and got success in all the experiments. In the first experiment, they collected forensic data remotely from the guest OS layer of cloud. Encase Servlets and FTK Agents are the remote programs, which were used to communicate and collect data. For the second experiment, they prepared an Eucalyptus cloud platform and collected data from the virtualization layer. In the third experiment, they tested the acquisition at the host operating system layer by Amazon's export feature. They found that though it is possible to export data from S3, it is not possible from EBS.

\section{Advantage}
\label{sec:advantage}
Though cloud forensics is a complicated process and imposes new challenges in digital forensic procedure, it offers some advantage over traditional computer forensics. Several researchers highlight the availability of computing environment through VM, which can be helpful to acquire the computing environment for forensic investigation \cite{reilly2011cloud,taylor2010digital}. We can use the VM image to use as a source of digital evidence. The computation and storage power of cloud computing can also boost up the investigation process \cite{sluskyforensic,taylor2010digital}. Cloud computing can reduce the time for data acquisition, data copying, transferimg and data cryptanalysis. Forensic image verification time will be reduced if a cloud application generates cryptographic checksum or hash. Ruan et al. highlighted some advantages of cloud forensics, such as, cost effectiveness, data abundance, overall robustness, scalability and flexibility, standards and policies, and forensics-as-a-service \cite{ruan2011cloud}. If the CSPs integrate forensic facilities in cloud environment, or they offer forensics-as-a-service to the customer by utilizing the immense computing power, then the customers do not need to implement any forensic schemes. In that way, cloud forensics will be cost effective for small and medium scale enterprise. Currently, Amazon replicates data in multiple zones to overcome the single point failure. In case of data deletion, this data abundance can be helpful to collect evidence. Amazon S3 automatically generates MD5 hash of an object when we store the object in      S3, which removes the need of external tools and reduces the time for generating hash. Amazon S3 also provides versioning support. From the version log, we can get some crucial information for investigation, such as, who accessed the data, and when, what was the requestor's IP, and what was the change in a specific version. Roussev et al. showed that for large-scale forensics analysis, cloud computing outperforms the tradition forensic computing technique \cite{roussev2009cloud}.

\section{Cloud computing usage in digital forensics}
\label{sec:usageindigital}
While cloud computing model often makes digital forensics difficult, the use of cloud computing technology can also facilitate the traditional digital forensic investigation procedure. Lin et al. proposed an RSA signature based scheme, where they showed how we can use the RSA signature scheme to safely transfer data from mobile devices to cloud storage \cite{lincloud}. It ensures the authenticity of data and thus helps in maintaining a trustworthy chain of custody in forensic investigations. By using RSA  signature  protocol, a  verifier  can  verify  the evidence in the court. They described the steps of uploading the digital evidences to the forensic data center preserving privacy and downloading for verification. In this process, the cloud computing center computes the RSA signature and send the signature to cloud storage center, which save the final output. The final output can later be downloaded to check the integrity of the data. They conducted an experiment of their method in both cloud and traditional environment and get better result in cloud. 

Buchanan et al.  proposed a cloud-based Digital Forensics Evaluation Test (D-FET) platform to measure the performance of the digital forensics tools \cite{buchanan2011performance}.  The quality metrics are: true-positives, false-positives, and operational performance (e.g., the speed of success, CPU utilization, and memory usage). They described how they set up the virtualization environment and how they ran their experiment there.

\section{Open problems}
\label{sec:problem}
In Section \ref{sec:solution}, we have seen that researchers have proposed several solutions to mitigate some challenges. Unfortunately, only a few of the proposed solutions have been tested with real world scenarios. Besides that, to the best of the authors' knowledge, CSPs have not adopted any of the proposed solution yet. There are a good number of open problems. Cloud management plane or API to get the necessary logs can decrease the dependence on CSP. However, as we do not have physical access, we still need to depend on CSP for various forensic data acquisition purposes, e.g., collecting temporary registry logs, identifying deleted files from hard disk, etc. Therefore, diminishing the dependence on CSP is still unsolved. 

Limited bandwidth is another critical issue. If the cloud storage is too high then bandwidth will be a challenge for time critical case. This issue has not been resolved yet. Several researchers have proposed secure data provenance to mitigate the chain of custody issue. However, no concrete work has been done yet, which can show how we can preserve the chain of custody by secure provenance. To mitigate the cross border issue, researchers have proposed global unity, but there is no guideline about how this will turn out into reality. Moreover, no solution has been proposed for crime scene reconstruction or presentation issues. Modifying the existing forensic tools, or creating new tools to cope up with cloud environment is another big issue that has not been resolved yet.

Several researchers also discussed some open problems of cloud forensics. About logging issue in cloud forensics, Marty proposed some open research topics in application level logging, which are: security visualization, forensic time-line analysis, log review, log correlation, and policy monitoring \cite{marty2011cloud}. Wolthusen identified another critical open problem, which is identifying the precise location and jurisdiction under which a certain datum lies \cite {wolthusen2009overcast}.

Table \ref{table:openproblem} provides an overview of the open problems in cloud forensics.
\begin{table}[t]
\begin{center}
	\renewcommand{\arraystretch}{1}
	\begin{tabular}{|p{0.45\textwidth} |}
	\hline
	\cellcolor[gray]{0} \textcolor{white}{Open Problems} \\ \hline 
	
	Overcome the dependence on the CSP \\ \hline
	Making proof of concept for cloud management plane and forensics-as-a-service \\ \hline
	Acquiring large volume of data remotely for time critical case\\ \hline
	Preserving chain of custody by secure data provenance \\ \hline
	Guideline and implementation of global unity to overcome the cross border issue \\ \hline
	Crime scene reconstruction in cloud environment \\ \hline
	Modifying the existing forensics tools to cope up with cloud paradigm \\ \hline
	Identifying the precise location and jurisdiction of certain datum \\ \hline
	Security visualization of logs \\ \hline
	Forensic time line analysis of logs \\ \hline
	Log review, log correlation and policy monitoring \\ \hline
	\end{tabular}
\end{center}
\caption{Open Problems of Cloud Forensics at a Glance}
\label{table:openproblem}
\end{table}

\section{Conclusion}
\label{sec:conclusion}
With the increasing use of cloud computing, there is an increasing emphasis on providing trustworthy cloud forensics schemes. Researchers have explored the challenges and proposed some solutions to mitigate the challenges. In this article, we have summarized the existing challenges and solutions of cloud forensics to answer the question -- \emph{Where does cloud forensics stand now?} Current research efforts suggest that cloud forensics is still in its infancy. There are numerous open problems that we have mentioned in Section \ref{sec:problem}. By analyzing the challenges and existing solutions, we argue that CSPs need to come forward to resolve most of the issues. There is very little to do from the customers' point of view other than application logging. All other solutions are dependent on CSPs and the policy makers. For forensics data acquisition, CSPs can shift their responsibility by providing robust API or management plane to acquire evidence. Legal issues also hinder the smooth execution of forensic investigation. We need a collaborative attempt from public and private organizations as well as research and academia to overcome this issue. Solving all the challenges of cloud forensics will clear the way for making a forensics-enabled cloud and allow more customers to take the advantages of cloud computing.

\bibliographystyle{IEEEtran}
\bibliography{cloudForensic}
\end{document}